\documentclass[acmsmall]{acmart}

\usepackage{dirtytalk}
\usepackage{graphicx}
\usepackage{balance}
\usepackage{lmodern} %
\usepackage{microtype} %
\usepackage[most]{tcolorbox}
\usepackage{makecell}
\usepackage{enumitem}
\usepackage{multirow}
\usepackage{listings}
\usepackage{array}
\usepackage{hyperref}

\newboolean{showcomments}
\setboolean{showcomments}{true}
\ifthenelse{\boolean{showcomments}}
  {\newcommand{\nb}[2]{
  \fbox{\bfseries\sffamily\scriptsize#1}
     {\sf\small$\blacktriangleright$\textit{\textcolor{red}{#2}}$\blacktriangleleft$}
   }
  }
  {\newcommand{\nb}[2]{}
   
  }

\newtcolorbox{facetbox}[1]{%
  title=\strut #1,
  fonttitle=\bfseries,
  colback=gray!3,       %
  colframe=gray!55,     %
  colbacktitle=gray!10, %
  coltitle=black,       %
  boxrule=0.5pt,
  arc=2pt,              %
  left=1em,right=1em,top=0.8em,bottom=0.8em,
  toptitle=0.4em,bottomtitle=0.4em,
  enhanced,
  breakable             %
}

\author{Felix Dobslaw}
\affiliation{%
  \institution{Mid Sweden University}
  \department{Department of Communication, Quality Management and Information Systems}
  \city{Östersund}
  \country{Sweden}
}
\email{felix.dobslaw@miun.se}

\author{Robert Feldt}
\affiliation{%
  \institution{Chalmers University of Technology}
  \department{Department of Computer Science and Engineering}
  \city{Gothenburg}
  \country{Sweden}
}
\email{robert.feldt@chalmers.se}

\author{Juyeon Yoon}
\author{Shin Yoo}
\affiliation{%
  \institution{KAIST}
  \department{School of Computing}
  \city{Daejeon}
  \country{Republic of Korea}
}
\email{{juyeon.yoon, shin.yoo}@kaist.ac.kr}

\begin{document}

\title{Challenges in Testing Large Language Model Based Software: A Faceted Taxonomy}

\begin{abstract}

Large Language Models (LLMs) and Multi-Agent LLMs (MALLMs) introduce non-determinism unlike traditional or machine learning software, requiring new approaches to verifying correctness beyond simple output comparisons or statistical accuracy over test datasets.
This paper presents a taxonomy for LLM test case design, informed by research literature and our experience. Each facet is exemplified, and we conduct an LLM-assisted analysis of six open-source testing frameworks, perform a sensitivity study of an agent-based system across different model configurations, and provide working examples contrasting atomic and aggregated test cases.
We identify key variation points that impact test correctness and highlight open challenges that the research, industry, and open-source communities must address as LLMs become integral to software systems.
Our taxonomy defines four facets of LLM test case design, addressing ambiguity in both inputs and outputs while establishing best practices. It distinguishes variability in goals, the system under test, and inputs, and introduces two key oracle types: atomic and aggregated. Our findings reveal that current tools treat test executions as isolated events, lack explicit aggregation mechanisms, and inadequately capture variability across model versions, configurations, and repeated runs. This highlights the need for viewing correctness as a distribution of outcomes rather than a binary property, requiring closer collaboration between academia and practitioners to establish mature, variability-aware testing methodologies.

\end{abstract}


\maketitle

\section{Introduction}

Large Language Models (LLMs) and Multi-Agent LLMs (MALLMs)~\cite{wang2024survey,he2024llm} are transforming software development—not just through their capabilities but also due to their inherent non-determinism. Unlike traditional systems, where variability arises from e.g. unreliable servers, stochastic sub-components, or real-time inaccuracies, LLMs exhibit fundamental unpredictability due to their construction, differences in model selection, configuration, and input variations at both syntactic and semantic levels. Conventional testing methods and approaches to oracle formulation and correctness~\cite{barr2014oracle} struggle in this setting. Based on our experience in designing, implementing, testing, and assessing several software systems\slash solutions either partly or fully built around LLMs~\cite{feldt2023towards,yoon2024intent,kaarre2024chatgpt,khoee2024gonogo,wang2025automating,Kang2023aa,Kang2024aa,Kang2024ay,Kang2025sn}, we propose a taxonomy for LLM test case design that captures key challenges and nuances in this emerging paradigm.

Recent research highlights the impact of input and output ambiguity in LLM-based applications. Subtle prompt variations can invert model responses, even under high-confidence settings~\cite{xiao2024automated}, and repeated queries--despite deterministic configurations (e.g., temperature = 0)--can produce inconsistent outputs~\cite{atil2024llm}. This variability raises concerns for replicability and necessitates advances in automated oracle design~\cite{molina2024test}.

Traditional software testing relies on deterministic oracles, but the probabilistic nature of LLMs challenges this assumption. While Barr et al.~\cite{barr2014oracle} introduced probabilistic oracles to handle non-determinism, their framework does not account for prompt-driven variability in LLMs and MALLMs. More broadly, ML testing requires a paradigm shift~\cite{braiek2020testing}, yet existing work does not explicitly treat variation as a first-class concern. The challenge is further compounded by LLMs' hybrid nature, where behavior emerges from a combination of code, model inference, and prompt engineering.

This divergence from traditional testing is evident in how correctness is defined. Existing paradigms--whether deterministic~\cite{barr2014oracle}, stochastic~\cite{grunske2008specification, grunske2009monitoring,barr2014oracle}, or ML-specific~\cite{riccio2020testing, dobslaw2023similarities}--struggle to address the multi-layered complexity of LLM-based systems. Unlike static ML models, where uncertainty stems from training data and inference, LLMs introduce an additional axis: the prompt itself, which acts as both input specification and behavioral modifier. Moreover, LLM performance may degrade post-deployment due to data shifts, yet monitoring mechanisms remain underdeveloped~\cite{braiek2020testing}. The disconnect between LLM testing tools and core ML or SE testing literature further underscores the need for specialized methodologies~\cite{hudson2024software}.

This article extends our previous work~\cite{dobslaw2025challenges} by presenting a faceted taxonomy for LLM test case design that organizes testing concerns across four dimensions: Software Under Test, Goal, Oracles, and Inputs. Central to this taxonomy is the distinction between \emph{atomic oracles}, which evaluate single executions under the assumption of determinism, and \emph{aggregated oracles}, which assess correctness across multiple runs to account for stochastic behavior. We validate the taxonomy through comprehensive empirical investigations that combine multiple methods: concrete examples demonstrating each facet, LLM-assisted analysis of six open-source testing frameworks (Opik, DeepEval, RAGAs, Promptfoo, Phoenix, and Giskard), a sensitivity study comparing different model configurations (GPT-3.5-turbo vs GPT-4) in the DroidAgent agent-based system, and working implementations contrasting atomic and aggregated test approaches.

Our findings reveal gaps across multiple dimensions of current testing tooling and practice. First, analysis of the six frameworks shows that tools predominantly treat each test execution as an isolated event, relying on atomic oracles that presume deterministic outcomes. Second, while some tools aggregate results at the dataset level (e.g., reporting pass rates), none provide explicit mechanisms for handling multiple evaluations of the same input within a single system configuration--ignoring the stochastic variability inherent to LLMs. Third, the sensitivity study demonstrates that even minor configuration changes (model selection, temperature settings) can dramatically alter system behavior, yet current tools lack systematic support for capturing and comparing such variability across model versions and configurations. These observations collectively challenge the assumption that correctness can be assessed through isolated executions and reveal that mature LLM testing requires viewing correctness as a distribution of outcomes rather than a binary property. This conceptual shift has implications not only for tool design and aggregation strategies but also for how testing responsibilities may need to extend beyond traditional software testers to include domain experts who possess the contextual knowledge necessary for validation in specialized domains.

The structure of the article is as follows. Section \ref{sec:background} offers background and motivates the need for a taxonomy and structured approach to the testing of LLM-based software. Section~\ref{sec:taxonomy} introduces our faceted taxonomy~\cite{usman2017taxonomies} for LLM test cases, categorizing key variation points that impact evaluation correctness, including the distinction between Atomic and Aggregated Oracles to address the identified gap. Unlike prior work--\cite{xiao2024automated} on adversarial robustness and \cite{hudson2024software} on broad LLM testing taxonomies--our framework organizes test case design across multiple facets, extending beyond specific testing foci and high-level categorization.
Section~\ref{sec:empirical} presents our empirical investigations, including tool mapping, LLM-assisted analysis, and sensitivity studies. Finally, Section~\ref{sec:2030} discusses implications for the field and identifies key challenges for ensuring correctness in LLM-based software toward 2030.

\section{Background and Motivation}
\label{sec:background}
The proposed taxonomy for LLM test cases is designed to support not only the traditional testing of finalized LLM-based systems but also to play a central role throughout the entire development, maintenance, and evolution phases. Given the inherent non-determinism, flexible behavior, and the vast and often unpredictable output space of LLMs, we believe that test cases will be an essential part of the development process of any LLM-based systems, far more so than for the traditional software systems. 

While the emphasis on the need for test cases throughout the development lifecycle strongly reminds us of Test-Drive Development (TDD), we argue that LLM-based systems would take on a more dynamic and iterative process than that of TDD as we know of. In conventional software development, TDD follows a relatively linear cycle: developers write a test case which declaratively specifies what the expected program behavior should be, implement code to pass the test, and maintain a stable, ``green'' test suite--unless the specification changes. However, in LLM-based systems, the development process is less deterministic. Developers often begin with only a vague understanding of ``what to test'', especially in domains with complex requirements (e.g., medical chatbots). The high degree of causal uncertainty--the inability to predict how an LLM will interpret and follow a given prompt--means that developers cannot statically reason about expected behaviors. Unlike traditional code, where the flow of execution can often be understood without running it, at least on lower levels, LLM behaviors must be explored through dynamic analysis, observing actual outputs to identify patterns, strengths, and failures.

This leads to a more fluid form of TDD, where both the Software Under Test (SUT) and the test cases themselves evolve together. Developers iteratively refine prompts, configurations, and even the underlying models, all while adapting and extending the test suite based on observed behaviors. Importantly, achieving a 100\% green test suite is often neither feasible nor the goal. Instead, the focus shifts to systematically reducing uncertainty and guiding the system toward acceptable behavior ranges.

Furthermore, during the design phase, it's not uncommon that developers experiment with different base LLMs to evaluate their suitability for specific components. While traditional testing phases typically assume a fixed system configuration, in LLM-based development, variability in models, prompts, and hyperparameters is a key part of the design process. This means that testing is not simply a validation tool but a primary mechanism for system exploration and refinement.

This interplay between model selection, prompt engineering, and test design underscores the importance of integrating testing as an ongoing, foundational activity. The taxonomy we propose below supports this need by offering a structured framework for test case design that remains valuable across all stages of a system's lifecycle--from early prototyping to deployment and long-term maintenance.

In this context, testing is not a final verification step but an integral driver of system development. This dynamic approach to TDD highlights that, for LLM-based systems, testing is even more critical than in traditional software development, where formal methods, type systems, and static analyses can partially guarantee behavior. In contrast, for LLM-centric systems, testing is often the primary means of ensuring reliability, safety, and functionality.

Traditional software testing relies on a specification that defines the expected behavior, interfaces, and constraints of the system under test (SUT). In LLM-based systems, this process becomes more complex due to the flexibility and ambiguity of natural language interactions. To manage this, developers often use \textit{prompt templates}—structured inputs with placeholders, such as:

\begin{tcolorbox}[colback=gray!10, colframe=black, sharp corners, boxrule=.3pt, width=\columnwidth]
\footnotesize
Write code to read \texttt{<FORMAT>} file in \texttt{<PROGRAMMING\_LANGUAGE>} with telling variable names and add code comments.
\end{tcolorbox}

These templates function like macros in traditional software, offering structured yet customizable input formats controlled by developers.

While prompt templates enable diverse inputs, they also complicate testing. LLMs typically generate free-text responses without strict adherence to requested formats, making automated evaluation challenging. To address this, \textit{invariant checks} assess both syntactic correctness (e.g., ensuring the output is valid Python code) and semantic quality (e.g., verifying meaningful documentation). However, LLM responses can be ambiguous or only partially fulfil a request, requiring correctness evaluations to account for deviations ranging from minor formatting errors to complex, context-dependent variations.

These challenges, along with strategies for handling them, are further explored in the Oracle section. The Inputs section discusses how test cases instantiate prompt templates with varying, concrete data for evaluation.

\section{Taxonomy for LLM Test Case Design}
\label{sec:taxonomy}

Testing LLMs presents unique challenges due to their non-deterministic behavior and the ambiguity inherent in both inputs (e.g., prompt phrasing and intent) and outputs (e.g., diverse, context-dependent responses). Beyond variability, practical concerns such as cost and granularity in test case design further complicate reliable evaluation. To address these challenges, we propose a structured taxonomy for LLM testing, organized around four core dimensions: \textbf{Software Under Test (SUT), Goal, Oracles, and Inputs}. Each of these core dimensions is further sub-divided into finer-grained sub-facets, capturing the specific factors that influence test case design and evaluation as well as points of variability between specific test case runs. This taxonomy provides a systematic framework for developing and refining test cases across the software lifecycle, emphasizing the need for adaptive and continuous testing strategies tailored to the dynamic nature of LLM-based systems. The order of these dimensions reflects their logical dependencies in test case design.

\begin{enumerate}
    \item \textbf{SUT} is presented first because it defines the underlying system or component to be evaluated. The SUT typically remains constant across multiple test cases, serving as the foundation upon which different testing scenarios are applied.

    \item The test case \textbf{Goal} comes next, as it specifies the unique objective of the test case and highlights the specific properties of it to be tested. While the SUT may remain unchanged, each test case targets a specific property, such as ensuring functional correctness or non-functional goals like safety, fairness, or robustness.

    \item \textbf{Oracles} are then defined for each property derived from the test case goal.

    \item \textbf{Inputs} are the final dimension, consisting of the datasets, user interactions, or synthetic prompts used to elicit responses from the SUT, which are then assessed using the defined oracles.
\end{enumerate}

This ordered structure can help ensure clarity in the design process. Starting with the SUT allows for reusable test infrastructures, while distinct goals enable focused evaluations. The oracles then provide concrete evaluation mechanisms for each property, and the inputs serve as the vehicle to generate diverse test executions. However, the taxonomy is not prescribing a particular order of analysis and developing and testing with LLMs is an inherently iterative process. Below we detail each dimension further and discuss its sub-facets.

\subsection{Case: Issue report classification}
We explain all four facets in support of a case SUT by name ClassifyIssueReport, which has the purpose to classify issue reports for issue tracking (such as on GitHub). Given an issue text, the SUT shall return exactly one label (for reasons of simplicity) - BUG, FEATURE, INVALID, or DUPLICATE. Invalid issue reports demand work without benefit \cite{laiq2025automatic}. In case of a duplicate, i.e. an issue that has been brought up already in an existing issue report, point to the matching existing issue. The classification then could lead to different actions, for example assigning bugs to an appropriate tester or developer, or adding a duplication note in the thread with reference to a duplicate issue and closing it.

Classifying reported issues has practical relevance and can be solved in many different ways in support of traditional software, machine learning, or LLMs. Using LLM's we could imagine simple solutions with a single model or agentic systems that collaborate over the entire projects issue reporting history to match potential duplicates.

\subsection{Software Under Test}

The system under test (SUT) refers to the software implementation being evaluated for correctness. In this context, we define the SUT as a software system that integrates one or more large language models (LLMs)--such as in multi-agent LLM architectures~\cite{yoon2024intent} (e.g., MALLM)--to achieve the expected behavior specified in its design.
To analyze variation both across test cases and within a single test execution, we break down the SUT in LLM-centric systems into the following key \textbf{sub-facets}.

The \textbf{Component} represents the specific functional unit or role within the system that is under test. In LLM-based systems, this often corresponds to a prompt (or prompt template), an instruction set, or an agent role. For instance, in a multi-agent setup, the component might be the \emph{Planner} agent (as in the multi-agent LLM system of \cite{yoon2024intent}) or a specific prompt guiding a data transformation task. However, our taxonomy remains agnostic to the level of testing: in a multi-agent LLM system, a targeted subsystem composed of multiple agents may itself serve as the SUT. The component facet thus defines the logical function under test, independent of the specific model(s) executing it.

The \textbf{Model(s)} refers to the LLM(s) implementing the component(s) during test execution. Different base models (e.g., \emph{Claude 3.5}, \emph{Deepseek R1}, \emph{GPT-4o}) can produce distinct behaviors even under identical prompts, making this a critical variation point. In MALLMs, each component may rely on a different model or a combination of models, requiring test cases to account for these permutations.

Each model operates under specific \textbf{Configurations} that directly influence its behavior. These include model parameters such as temperature, top-k/top-p sampling, and maximum token limits, as well as system-level settings like API rate limits and external tool integrations. Configurations may vary per model or be shared across components, introducing additional variability that must be managed in test design.

In the context of a specific test case, the SUT is a concrete instance with fixed choices for the Component, Model(s), and Configuration(s). However, our taxonomy explicitly distinguishes between fixed and variable elements across test runs. Any modification--such as swapping the underlying LLM, adjusting a temperature setting, or altering the role’s prompt--creates a new SUT instance that may behave differently under the same input.

This approach contrasts with traditional software testing, where the SUT is typically treated as fixed and configuration changes are minor or peripheral. In LLM-based systems, even small variations can significantly impact behavior, making them a \textbf{deliberate factor in test case design rather than an unintended side effect}. The complexity increases further in hybrid systems combining LLMs with conventional code, where changes in either can affect overall system behavior.

\begin{facetbox}{SUT Example}
\textbf{Components.} The \texttt{ClassifyIssueReport} SUT is a combination of a duplication checking function \texttt{DuplicationFinder} and an LLM call. For a given issue report, the function returns either a matching issue id - to then return DUPLICATE - or null. In the latter case the LLM is invoked.

\textbf{Model and Configuration.} One LLM is used only when the duplication check returns null: Mistral-7B-v0.1. The \texttt{DuplicateFinder} tool is used in version 0.3. The LLM is called with a fixed prompt defining BUG, FEATURE, and INVALID and instructing to answer with \textit{exactly} one of them. Result variability is limited using temperature=0.0, top\_p=1.0, n=1, and max\_tokens=16.
\end{facetbox}

\subsection{Goal}

The goal defines the high-level objective of a test case, which is then refined into specific, measurable sub-goals called \textbf{Properties}--the concrete conditions the SUT must satisfy. For instance, if the goal is to \textit{ensure the safety of LLM outputs}, relevant properties might include that the LLM must not generate offensive language (Property 1) and must not encourage harmful behavior (Property 2). These properties form the basis for designing oracles that evaluate whether the SUT meets the defined goals.

\begin{facetbox}{Goal Examples}
\textbf{G1 Decision validity.} Given one issue report, the system should return exactly one label from {BUG, FEATURE, INVALID, DUPLICATE}.
\begin{itemize}
\item \textbf{P1.1} Single-label: the output contains exactly one label.
\item \textbf{P1.2} Membership: the label is strictly one of {BUG, FEATURE, INVALID, DUPLICATE}.
\item \textbf{P1.3} Format discipline: no extra content beyond the label.
\end{itemize}

\textbf{G2 Duplicate alignment.} When the duplication function provides a matching id, the system should surface a DUPLICATE decision that references that id; when it provides no match, the system should decide among BUG, FEATURE, INVALID.
\begin{itemize}
\item \textbf{P2.1} Tool consistency: if the duplication function returns an id, the decision is DUPLICATE and the same id is echoed.
\item \textbf{P2.2} Non-duplicate discipline: if the duplication function returns null, the decision is not DUPLICATE.
\item \textbf{P2.3} ID provenance: any reported duplicate id must come from the duplication function.
\end{itemize}

\textbf{G3 Consistent behavior.} At fixed prompt/model/settings, the system should produce the same decision for the same input text.
\begin{itemize}
\item \textbf{P3.1} Run-to-run stability: at fixed settings, the same input yields the same label.
\item \textbf{P3.2} Minor-text robustness: light, meaning-preserving paraphrases yield the same label.
\item \textbf{P3.3} Prompt-format robustness: benign formatting changes (e.g., whitespace, line breaks) do not change the label.
\end{itemize}
\end{facetbox}

\subsection{Oracles}
Oracles determine whether the SUT meets the defined properties we have decided upon given the test case goal. We make a novel distinction between the two levels of \textit{Atomic} and \textit{Aggregated} Oracles, and emphasize the latter’s importance in LLM-based testing.

An \textbf{Atomic Oracle} evaluates correctness based on a \textit{single} test execution, much like in traditional software testing, where outputs are compared against predefined criteria. Atomic oracles may use strict deterministic checks, such as equality comparisons, regular expressions, or structural matching--effective for predictable outputs like numerical computations. However, LLM-generated responses introduce inherent variability, frequently making such rigid criteria insufficient. To address this, Atomic Oracles can also incorporate probabilistic or heuristic-based evaluations, including e.g. semantic similarity measures, rule-based checks for required keywords, and human-in-the-loop validation for subjective assessments. 
While these alternatives can be useful, they often will still fail to fully account for LLMs' non-deterministic nature, even when the LLMs are configured for determinism (e.g., temperature = 0)~\cite{atil2024llm}.

An \textbf{Aggregated Oracle} mitigates this limitation by evaluating correctness across \textit{multiple} test runs under the same conditions. Since LLM outputs vary for the same input, multiple samples can help establish consistency, and correctness is better assessed through statistical aggregation rather than single-instance evaluations. These oracles rely on Atomic Oracles for individual test executions and then apply an aggregation function to derive a final verdict. Aggregation strategies include measuring result variance, majority voting, confidence-weighted scoring, or reference-based comparisons against paraphrased expected outputs.

This distinction has major implications for LLM-based testing. Traditional methods struggle with non-deterministic outputs, necessitating batch testing frameworks that repeatedly execute test cases to compute aggregated verdicts. The choice of aggregation strategy depends on the application: structured data extraction may allow stricter Atomic Oracles, while creative text generation requires a more \textit{nuanced} Aggregated Oracle approach.

By bridging deterministic software testing with LLM stochasticity, Aggregated Oracles provide a \textit{robust} and \textit{realistic} methodology for evaluating correctness in an inherently variable domain.

\begin{facetbox}{Oracle Examples}
\textbf{For G1 Decision validity.}
\begin{itemize}
\item \textbf{O1 (P1.1--P1.3)} Per-case check: read the output; confirm it has exactly one label, that the label is in {BUG, FEATURE, INVALID, DUPLICATE}, and that nothing extra is included.
\item \textbf{O1-metric} Dataset view: share of cases that pass O1.
\end{itemize}

\textbf{For G2 Duplicate alignment.}
\begin{itemize}
\item \textbf{O2 (P2.1)} Per-case check when the tool returns an id: label is \texttt{DUPLICATE} and the same id is returned.
\item \textbf{O3 (P2.2)} Per-case check when the tool returns \texttt{null}: label is not \texttt{DUPLICATE}.
\item \textbf{O2/O3-metric} Dataset view: share of cases that follow the tool result (report separately for tool=id and tool=null).
\end{itemize}

\textbf{For G3 Consistent behavior.}
\begin{itemize}
\item \textbf{O4 (P3.1)} Metric: repeatability — same label across re-runs with fixed settings.
\item \textbf{O5 (P3.2)} Metric: paraphrase agreement — same label across light rewordings.
\item \textbf{O6 (P3.3)} Metric: format agreement — same label across harmless formatting changes.
\end{itemize}
\end{facetbox}

\subsection{Inputs}

Inputs are the concrete prompts that drive SUT execution, forming the final dimension of the taxonomy. They typically populate placeholders in prompt templates with specific dataset values, simulating real user interactions or system executions. However, in systems allowing multi-turn dialogues, having reasoning agents, or stateful architectures, later inputs often depend on earlier outputs, which can make static substitutions insufficient.

In addition to specifying the specific data values or how to derive sequences of inputs, the Inputs dimension can also list explicit variability operators to assess the SUT's robustness. These fall into two categories: \textit{syntactic variations}, which modify formatting or wording while preserving meaning, and \textit{semantic variations}, which shift the meaning to explore different behaviors. Such variations apply at both the prompt template and input levels, expanding test coverage and blurring the distinction between prompt templates as input (from the LLM’s perspective) and as part of the SUT (as a development artefact).

This layered approach allows comprehensive probing of the SUT, generating diverse outputs for evaluation by oracles. Unlike traditional software testing, where inputs map directly to expected outputs, LLM interactions introduce ambiguity--user inputs may refine, complement, or complete a prompt, complicating clear test case boundaries.

Systematically addressing input variability involves creating comprehensive input coverage strategies that incorporate both syntactical and semantical dimensions. This approach not only ensures higher reliability and robustness of LLM-based software but also supports clearer identification of failure points related to input ambiguities.

\begin{facetbox}{Inputs Example (Adequacy)}
\textbf{What we check.} We declare the input space \emph{adequately covered} when the following simple, countable conditions hold:
\begin{itemize}
\item \textbf{Class balance.} Each base item class (BUG, FEATURE, INVALID, DUPLICATE) has at least 50 items.
\item \textbf{Syntactic variation.} For every base item, generate exactly 3 meaning-preserving variants: (S1) whitespace/line-break change, (S2) punctuation/case tweak, (S3) benign formatting edit.
\item \textbf{Semantic variation (boundary).} For 20\% of items per class, add 2 near-miss paraphrases that push a neighboring class boundary (e.g., BUG$\leftrightarrow$FEATURE).
\end{itemize}

\textbf{How we record it.} Each row stores \texttt{base\_id}, \texttt{class}, \texttt{variant\_type} $\in$ \{BASE, S1, S2, S3, SEM1, SEM2\}. A simple adequacy report lists per-class counts and per-operator coverage:
\begin{itemize}
\item \textbf{Per-class coverage:} target met if $\geq$ 50 BASE items and $\geq$ 150 syntactic variants (3 per BASE) per class.
\item \textbf{Boundary coverage:} target met if $\geq$ 10%
\item \textbf{Stop rule:} declare input adequacy when all targets are met; otherwise, generate the missing variants only.
\end{itemize}
\end{facetbox}

\section{Empirical Investigations}
\label{sec:empirical}

To demonstrate and test the practical value of our taxonomy, we conducted three complementary empirical investigations. Together, they examine how the taxonomy can guide both human and automated tool evaluation and how selected aspects of it manifest in practice. The first two studies focus on testing tools: Section \ref{sec:mapping} presents a manual mapping of six open-source frameworks, while Section \ref{sec:prompt} extends this analysis by turning the taxonomy into a detailed checklist that also serves as a structured prompt for LLM-based evaluation. This dual approach allowed us to assess how well the taxonomy captures existing practices and to explore whether LLMs can assist in identifying tool capabilities and gaps not yet represented in the taxonomy. The third study, in Section \ref{sec:sensitivity}, shifts focus from tools to systems under test, applying a sensitivity analysis to a multi-agent Android testing system \cite{yoon2024intent} to illustrate how SUT variability appears across different LLM models and leads to output variability that motivates aggregated oracles.  All supplementary materials, including tool documentation, analysis prompts, detailed results, and complete action logs, are available online\footnote{\url{https://doi.org/10.5281/zenodo.17393106}}.

\subsection{Manual Taxonomy-Based Tool Evaluation}
\label{sec:mapping}

Here we analyze how each facet of our taxonomy is addressed by existing tools that represent the state of practice in LLM application testing. Several open-source~\cite{deepeval, promptfoo, pareaAI, trulens, opik, giskard} and hosted~\cite{wandb, langsmith} platforms have emerged to support this domain. This paper focuses on a selection of widely used open-source frameworks as representative examples.  

We identified relevant GitHub repositories by searching for ``LLM Testing'' and ``LLM Evaluation'', selecting frameworks that specifically target LLM applications (i.e., systems combining LLMs with prompt templates, optionally enhanced with retrieval mechanisms or tool-calling capabilities) and provide dedicated testing interfaces. The six selected projects, chosen based on GitHub popularity (stars) and distinct testing features, are Opik (14.4k), DeepEval~\cite{deepeval} (11.4k stars), RAGAs (11k stars), Promptfoo~\cite{promptfoo} (8.6k stars), Phoenix (7.1k), and Giskard~\cite{giskard} (4.9k stars) at the time of writing.

\newcolumntype{P}[1]{>{\centering\arraybackslash}m{#1}}

\begin{table*}[h]
\small
    \centering
    \caption{Mapping of a selection of open-source testing frameworks to the taxonomy}
    \label{tab:toolings}
    \resizebox{\textwidth}{!}{%
    \begin{tabular}{|P{1.4cm}|P{3.7cm}|P{3.7cm}|P{3.7cm}|P{3.7cm}|}
        \hline
        \textbf{Facet} & \textbf{SUT} & \textbf{Goal} & \textbf{Oracles} & \textbf{Inputs} \\
        \hline
        \textbf{Opik} 
        & The tool supports visualization features for comparing different models and prompt versions.
        & Testing goal can be set with one or more properties with varying number of inputs (dataset).
        & Oracles are implemented using deterministic, context relevance, and LLM-as-a-judge metrics.
        & Inputs with semantic variations can be synthesized (expanded from existing input examples) with an separate LLM agent and with custom variation instructions from users.
        \\
        \hline
        \textbf{DeepEval} 
        & The tool supports visualization features for comparing different models and prompt versions.
        & Testing goal can be set with one or more properties for each test input.
        & Oracles are implemented using deterministic and model-based metrics, including conversational/multi-modal support.
        & Inputs with semantic variations can be synthesized with an separate LLM agent and with various resources (e.g., documents).
        \\
        \hline
        \textbf{RAGAs} 
        & The tool supports organizing responses from different models and prompt versions, while keeping the visualization features private
        & Testing goal can be set with one or more properties with varying number of inputs (dataset).
        & Oracles are implemented as metrics measuring context relevance, tool call accuracy, and similarity with reference answers either with llms or deterministic calculation.
        & Inputs, specifically the queries for the retrieval-augmented systems, can be synthesized using a knowledge graph extracted from documents to ensure semantic diversity.
        \\
        \hline
        \textbf{Promptfoo} 
        & The tool supports visualization features for comparing different models and prompt versions.
        & Testing goal can be set with one or more properties with varying number of inputs.
        & Oracles are implemented using deterministic and model-based metrics, allowing weighted metric combinations.
        & Inputs with semantic variations can be synthesized with an separate LLM agent.
        \\
        \hline
        \textbf{Phoenix} 
        & The tool supports visualization features for comparing different models and prompt versions.
        & Testing goal can be set with one or more properties with varying number of inputs.
        & Oracles are implemented using deterministic and LLM-as-a-judge metrics.
        & Inputs can be synthesized with a custom prompt and an LLM.
        \\
        \hline
        \textbf{Giskard} 
        & The tool separates visualization and result comparison features into an external enterprise platform.
        & Testing goal can be defined as one or more properties across a set of inputs.
        & Oracles are typically implemented as dataset-wise scores via built-in or custom metric functions across multiple inputs.
        & Inputs can be transformed using various predefined syntactic transformation methods.
        \\
        \hline
    \end{tabular}
    } %
\end{table*}

\subsubsection*{SUT}
All tools accommodate variability across the three SUT sub-facets, though their capabilities differ slightly. Each allows users to specify the \textit{Component} under test (e.g., a prompt template or agent), select the \textit{Model(s)} (e.g., GPT-4o, Claude 3.7 Sonnet), and adjust \textit{Configurations} (e.g., temperature, token limits) that influence runtime behavior.  
Notably, Promptfoo enables ``unfolded'' comparisons, allowing testers to assess differences between prompt versions, model choices, or configuration settings. Similarly, Opik and DeepEval include visualization tools but primarily focus on identifying regressions through aggregated metrics (e.g., the proportion of passing test cases within a dataset). While Giskard's open-source SDK does not directly support cross-version SUT comparisons, this functionality is available through integration with an an external enterprise platform~\cite{giskard-hub-ui}.  

\textbf{Assessment:} Each tool, to some degree, recognizes that developers frequently re-evaluate multiple implementations throughout development. They support continuous testing by enabling iterative modifications to components, models, and configurations, while also facilitating some level of structured comparisons of test results.

\subsubsection*{Goal}

An ideal test case is defined with a clear and specific testing goal, typically composed of one or more relevant properties. In most frameworks, a test case consists of a single input paired with an expected output, evaluated against one or more property-checking metrics. Giskard places greater emphasis on dataset-level scores and uniquely offers fine-grained controls for selected properties in each test run, though such functionality is supported through the external enterprise platform. Certain LLM testing sub-goals---such as assessing robustness against prompt injection attacks---are best evaluated using a set of targeted inputs combined with specific properties of interest. While other tools can execute multiple test cases in a single evaluation, they lack explicit support for linking evaluations to a coherent test goal across different properties. Conversly, Giskard lacks the flexibility to configure individual test cases separately when needed.

Together, these frameworks fall short of balancing flexibility (checking properties at the individual input level) with coherence (grouping relevant inputs under a common evaluation goal).  

\textbf{Assessment:} While these tools offer various property-checking mechanisms, they struggle with goal variability and representability--the ability to flexibly define and assess test goals at different levels of granularity. Although properties serve as concrete criteria derived from test goals, current frameworks do not explicitly model this distinction or integrate it into evaluation workflows. Crucially, no standardized approach differentiates goals from properties at the test case level.

\subsubsection*{Oracles}
The tools support a range of oracles, often referred to as ``metrics'' or ``assertions,'' including deterministic checks (e.g., substring matching) and heuristic, model-based evaluations. Users can define custom metrics using tailored prompts for a separate LLM instance (\textit{LLM-as-a-judge})~\cite{gu2024survey, zheng2023judging} or through code-based implementations. Web interfaces enable manual review and labeling, but human-in-the-loop validation remains underdeveloped. While tools allow human-provided labels, they lack mechanisms for iterative collaboration between humans and model evaluators.  

Evaluations typically yield either a continuous score (e.g., similarity to a ground-truth output on a 0-1 scale) or a binary pass/fail result, with continuous scores often thresholded into binary outcomes. Giskard mainly operates at the dataset level, aggregating individual assessments (e.g., reporting a 70\% pass rate for politeness). However, LLM outputs can vary across repeated runs, and current tools lack aggregation methods to handle multiple evaluations of the same input within the same system under test (SUT). Instead, each run is either treated as an independent outcome, or simply averaged (Phoenix), ignoring stochastic variability.

We argue for aggregated oracles that account for repeated evaluations, using different aggregation strategies depending on the property under test and the source of variation. For example, a strict requirement such as ``the response MUST follow the specified JSON format'' should pass only if no violations occur across multiple runs. Conversely, a softer requirement like ``the response should be polite'' could be satisfied if the majority of runs meet the criterion.  

\textbf{Assessment:} While these tools offer diverse oracle definitions, model-based heuristic oracles remain unstable, and deterministic oracles lack expressiveness. Existing dataset-level aggregation is limited, and current tools fail to systematically capture variability across repeated runs. A principled approach to aggregated oracles is needed, along with flexible aggregation schemes (e.g., majority vote, strict pass, confidence-based thresholds) tailored to different evaluation properties.

\subsubsection*{Inputs}
Developers often struggle with a lack of data or inputs when testing LLM applications. Existing tools mitigate this by generating synthetic inputs using separate LLM instances or modifying existing inputs with predefined operators. These transformations span both \textit{syntactic} (e.g., punctuation removal, case conversion, language change) and \textit{semantic} (e.g., content diversification while preserving format) variations. Many tools also support adversarial input generation for security and safety testing, aligning with known attack methods~\cite{owasp}, such as injecting jail-breaking prompts or embedding encrypted forbidden content to bypass security filters.  

Giskard extends syntactic transformations with metamorphic testing, ensuring that minor syntactic changes do not significantly alter outputs. However, these input generation methods lack well-defined quantitative objectives, making it difficult to ensure comprehensive test coverage. This raises key questions: How can we determine whether available inputs are sufficient for thorough testing? How can we optimize regression testing to balance cost and rigor?

\textbf{Assessment:} A major limitation in existing tools is the absence of clear test adequacy criteria for input variability. While they provide input synthesis and data collection features, they do not systematically assess whether variations sufficiently test SUT robustness.

Overall, the manual mapping provided a structured view of how current testing frameworks address the key facets of our taxonomy. Yet, performing such detailed analyses manually is time-consuming and prone to subjective interpretation. To explore whether this process can be accelerated and standardized, we next examined whether large language models themselves can assist in evaluating testing tools.

\subsection{LLM-based Tool Evaluation Through a Detailed Checklist\slash Prompt}
\label{sec:prompt}

Building directly on the manual analysis, we transformed the taxonomy into a detailed checklist that operationalizes each facet as a set of concrete evaluation questions. This checklist serves a dual purpose: it can guide human analysts toward systematic coverage and also functions as a structured prompt for LLM-based evaluation. Using it, we investigated whether an LLM could analyze tool documentation with sufficient precision to replicate or complement human judgment.

To test this approach, we applied it to the same six tools by consolidating their documentation and using the checklist as a detailed prompt to Claude Sonnet 4.5. This method produced a reusable artifact for standardized tool comparison, validated that the taxonomy captures relevant dimensions of real-world tools, and revealed both current strengths and gaps. The complete checklist and the individual analysis results for each tool, together with the consolidated LLM-based summaries, are available in the online replication package; below we summarize the checklist’s structure and key insights.

The checklist operationalizes the taxonomy’s four core facets through targeted sub-questions that probe concrete tool capabilities. Each sub-capability is rated on a four-point scale (not supported, limited, partial, strong) and grounded in direct evidence from documentation. The framework also includes cross-cutting criteria such as reproducibility, cost control, security, and ecosystem integration, as well as a ``Beyond the Taxonomy'' section that captures emerging or unexpected features. This section also serves as an indirect evaluation of the taxonomy itself, revealing tool capabilities or dimensions that are not yet well represented within its current structure.

Applied as an LLM prompt, the checklist yielded coherent, high-level summaries and consistent facet-level characterizations. However, the model often misjudged finer-grained criteria, mistaking superficial mentions for true support---for instance, labeling any reference to repeated execution as ``partial support'' for aggregated oracles even when statistical analysis was absent. Thus, while LLMs can accelerate preliminary assessments and highlight relevant documentation, human interpretation remains essential for judging depth and context. Overall, the checklist offers a structured, reusable foundation for both automated and manual evaluation of LLM testing tools.

\subsubsection{LLM-based analysis of Current Tool Landscape}

Applying the checklist as a prompt with Claude Sonnet 4.5 to analyse the documentations of the six tools revealed consistent patterns and recurring gaps in how current frameworks support LLM testing.

Across all tools, atomic oracles are the most mature capability. Each framework provides extensive mechanisms for single-run evaluations, ranging from deterministic checks to LLM-as-judge configurations. In contrast, aggregated oracles, which are essential for handling non-determinism, are almost entirely absent. None of the tools offer statistical aggregation, confidence intervals, or variance-based analysis. Where reproducibility features exist, they only fix random seeds rather than addressing inherent stochastic variability. This lack of aggregation mechanisms limits the reliability of test verdicts and highlights a core area where future tools must advance.

Similarly, while input generation is widely supported, coverage and adequacy assessment remain underdeveloped. Most tools can synthesize datasets and adversarial prompts but provide little guidance on whether the explored input space is sufficient. Traditional notions such as code coverage lack clear analogues for high-dimensional, semantically rich inputs, leaving practitioners uncertain about when testing is comprehensive.

Finally, the ``Beyond the Taxonomy'' analyses exposed several dimensions that extend or challenge the current framework. The development–production continuum emerged as a missing axis, with many tools blurring the boundary between testing, monitoring, and deployment. While our taxonomy focuses on test design during development it would be interesting future work to evaluate if a broader view could extend usefulness further. Domain-specific testing paradigms, such as Ragas’s knowledge graph–based evaluation and DeepEval’s ConversationSimulator, suggest that specialized approaches may warrant explicit recognition. However, we argue that the core of the taxonomy should be generally applicable and thus not domain-specific. In addition, adoption and usability factors---including documentation quality, community support, and ease of integration---strongly influence tool uptake yet fall outside the taxonomy. Together, these insights show that the ``Beyond the Taxonomy'' section not only captures emerging innovation but also serves as a mechanism for evaluating and evolving the taxonomy itself.

\begin{table}[!ht]\centering
\caption{Repeated runs of actor agent from DroidAgent using GPT-3.5 and GPT-4o\label{tab:variability}}
\resizebox{\textwidth}{!}{
\begin{tabular}{lrrrrrl}\toprule
Model & Run & Succeed  &\# Actions  & \# Effective & \# Redundant & Cause of Failure \\\midrule
\multirow{10}{*}{GPT-3.5} & 1  & \text{X} & 20 & 0 & 0  & Failed to proceed from permission popup \\
                          & 2  & \text{X} & 12 & 5 & 7  & Failed to input a valid deck name (was an empty string) \\
                          & 3  & \text{X} & 20 & 3 & 17 & Failed to pass the permission popup (mistakenly clicks back buttons) \\
                          & 4  & \text{X} & 20 & 2 & 18 & Failed to pass the permission popup \\
                          & 5  & \text{X} & 16 & 5 & 11 & Failed to input a valid deck name (was an empty string) \\
                          & 6  & \text{O} & 20 & 7 & 13 & - \\
                          & 7  & \text{X} & 20 & 2 & 18 & Failed to pass the permission popup \\
                          & 8  & \text{X} & 20 & 1 & 19 & Failed to pass the permission popup \\
                          & 9  & \text{X} & 20 & 0 & 20 & Failed to pass the permission popup \\
                          & 10 & \text{X} & 20 & 1 & 19 & Failed to pass the permission popup \\ \midrule
\multirow{10}{*}{GPT-4o}  & 1  & \text{O} & 8  & 8 & 0  & - \\
                          & 2  & \text{X} & 20 & 3 & 17 & Mistakenly got into the app info page from the permission popup \\
                          & 3  & \text{O} & 7  & 7 & 0  & - \\
                          & 4  & \text{O} & 7  & 7 & 0  & - \\
                          & 5  & \text{O} & 8  & 8 & 0  & - \\
                          & 6  & \text{O} & 7  & 7 & 0  & - \\
                          & 7  & \text{O} & 7  & 7 & 0  & - \\
                          & 8  & \text{O} & 7  & 7 & 0  & - \\
                          & 9  & \text{O} & 8  & 8 & 0  & - \\
                          & 10 & \text{O} & 7  & 7 & 0  & - \\
\bottomrule
\end{tabular}
} %
\end{table}

\subsection{LLM sensitivity analysis}
\label{sec:sensitivity}

Our third investigation shifts attention from testing tools to the systems under test themselves. By executing the same LLM-based agent multiple times—each time implemented with a different underlying model—and observing the resulting outputs, we explore how the taxonomy’s SUT variability and Oracle facets appear in practice. This small empirical study highlights concrete needs for future LLM testing frameworks. Table~\ref{tab:variability} summarizes ten repeated runs of the actor agent in DroidAgent~\cite{yoon2024intent}, an autonomous GUI testing system for Android applications. In each run, the agent receives the prompt \texttt{Create a new flashcard deck named `Computer Science'} and performs the corresponding actions in the open-source app AnkiDroid. We compare results using two underlying models, GPT-3.5 and GPT-4o, to analyze behavioral differences across SUT configurations. The table’s columns---\# of actions, \# effective, and \# redundant---indicate, respectively, the total actions taken, those that produced meaningful GUI changes, and those that did not. Below, we first examine the variability in outcomes, motivating the use of aggregated oracles, and then discuss configuration variability across models.

\subsubsection{Nondeterministic Outputs and Usage of Aggregated Oracle}

Results in Table~\ref{tab:variability} immediately show the variability of LLM outputs. The functional variability concerns whether the actor has successfully executed the given scenario. Assuming that we can verify whether the agent succeeded or not based on the final status of the GUI, the following (simplified) test case can check the result of the agent behaviour as well as whether it succeeded within the given budget (i.e., the number of GUI actions):

\lstset{
    language=Python,
    basicstyle=\ttfamily\footnotesize, %
    keywordstyle=\color{blue}, %
    commentstyle=\color{green!50!black}, %
    stringstyle=\color{red}, %
    showstringspaces=false, %
    backgroundcolor=\color{gray!10}, %
    frame=single %
}

\begin{lstlisting}
def test_droidagent_atomic_deterministic_result():
  with open("scenario.txt", "r") as s:
      test_scenario = s.readlines()
      agent = DroidAgent(model="gpt-4o")
      agent.execute(test_scenario)
      
      assert agent.get_current_screen().contains_text("Computer Science")
      assert agent.get_num_actions() < 20
\end{lstlisting}

However, there also exists non-functional variability. That is, even succesful actions can vary in concrete details while being semantically equivalent. For example, consider the following two actions steps generated by GPT-4o:

\begin{minipage}[t]{0.48\linewidth}
\begin{lstlisting}[language={}, basicstyle=\ttfamily\tiny, caption=Action log with length 7, breaklines=true, frame=single, backgroundcolor=\color{gray!5}]
[step_0] (from the permission dialog page) I tapped the "Allow" button to grant notification permissions to AnkiDroid.
    > adb shell input tap [540, 1314]

[step_1] (from the AnkiDroid permission dialog page) I tapped the "OK" button to dismiss the storage permission request dialog.
    > adb shell input tap [880, 1700]

[step_2] (from the permission dialog page) I tapped the "Allow" button to grant storage permissions to AnkiDroid.
    > adb shell input tap [540, 1345]

[step_3] (from the main deck selection page) I tapped the "Add" button to initiate the creation of a new deck.
    > adb shell input tap [941, 2200]

[step_4] (from the add deck page) I tapped the "Create deck" button to start the process of naming a new deck.
    > adb shell input tap [865, 1615]

[step_5] (from the create deck page) I tapped the text field to focus it and input "Computer Science" as the new deck name.
    > adb shell input tap [540, 1233]
    > adb shell input text ['Computer Science']

[step_6] (from the create deck page) I tapped the "OK" button to confirm the creation of the new deck named "Computer Science".
    > adb shell input tap [880, 1400]




.
\end{lstlisting}
\end{minipage}
\hfill
\begin{minipage}[t]{0.47\linewidth}
\begin{lstlisting}[language={}, basicstyle=\ttfamily\tiny, caption=Action log with length 8, breaklines=true, frame=single, backgroundcolor=\color{gray!5}]
[step_0] (from permission prompt page) I allowed AnkiDroid to send notifications to dismiss the permission prompt.
    > adb shell input tap [540, 1315]

[step_1] (from welcome screen) I tapped the "OK" button to dismiss the welcome message and proceed with the setup.
    > adb shell input tap [880, 1691]

[step_2] (from permission prompt page) I tapped the "Allow" button to grant AnkiDroid access to files.
    > adb shell input tap [540, 1346]

[step_3] (from deck picker page) I tapped the "Add" button to proceed to the next step of creating a new flashcard deck.
    > adb shell input tap [940, 2200]

[step_4] (from deck picker page) I tapped the "Create deck" button to start creating a new flashcard deck.
    > adb shell input tap [865, 1615]

[step_5] (from create deck page) I tapped on the EditText field to focus on it for entering the deck name.
    > adb shell input tap [540, 1233]

[step_6] (from create deck page) I entered the text 'Computer Science' into the focused EditText field to name the new flashcard deck.
    > adb shell input text ['Computer Science']

[step_7] (from create deck page) I tapped the "OK" button to confirm and create the new flashcard deck named 'Computer Science'.
    > adb shell input tap [880, 1397]
\end{lstlisting}
\end{minipage}

In case we want to check the agent behaviour against a specific human reference, such non-functional variability would prevent us from using simple Boolean predicates. Instead, we need to capture the semantic contents of the generated output (in this case, the action logs) and compare its similarity to the reference using a threshold:

\begin{lstlisting}
def test_droidagent_atomic_semantic_realistic():
  with open("scenario.txt", "r") as s, open("reference_actions.txt", "r") as r:
      test_scenario = s.readlines()
      agent = DroidAgent(model="gpt-4o")
      agent.execute(test_scenario)
      actions = agent.get_action_log()

      human_reference = r.readlines()

      similarity = measure_similarity(actions, human_reference)
      assert similarity > threshold
\end{lstlisting}

Some of the properties may require other LLMs for checking due to their abstract nature. Suppose we have to use another LLM as a judge to determine whether a given action log contains any redundant action step or now. The simplified test case would look like the following:

\begin{lstlisting}
def test_droidagent_atomic_deterministic_redundancy():
  with open("scenario.txt", "r") as s:
      test_scenario = s.readlines()
      agent = DroidAgent(model="gpt-4o")
      agent.execute(test_scenario)
      
      actions = agent.get_action_log()

      prompt = f"""An Android GUI testing agent tried to execute the following
      test scenario: {test_scenario}. It has executed the following actions: 
      {actions}.

      Your task is to assess whether the actions taken by the agent are
      efficient and do not contain any redundant steps. Answer yes if this is
      the case, otherwise no. Do not add any explanation."""
      
      assessment = ollama.generate(model="gpt-4o", prompt=prompt)
      assert "yes" in assessment
\end{lstlisting}

Given the variability across ten different runs shown in Table~\ref{tab:variability}, we argue that any of the test cases shown above should be applicable to a number of runs, the results of which are then aggregated. For example, we can rewrite the functional test cases above for repeated runs as follows:

\begin{lstlisting}
def test_droidagent_aggregated_deterministic_result():
  with open("scenario.txt", "r") as s:
    test_scenario = s.readlines()
    agent = DroidAgent(model="gpt-4o")
    for i in range(10):
      agent.execute(test_scenario)
      num_passes += 1 if \
        agent.get_current_screen().contains_text("Computer Science") and \ 
        agent.get_num_actions() < 20
  assert num_passes / num_scenarios > threshold  
\end{lstlisting}

While here we illustrate the aggregation over multiple runs of the same prompt, it is also possible to aggregate model behaviour across different prompts: for example, we can evaluate the actor agent using aggregated results from multiple scenarios. However, we note that support for aggregating repeated runs and customising the aggregation methods for domain needs is currently lacking in most frameworks.

\subsubsection{SUT Variability: Model Versions}

Another observation from Table~\ref{tab:variability} is that GPT-4o achieves a much higher success rate compared to GPT-3.5. While this is due to the advances in the foundation model and thus welcome, it does also raise the need for regression testing for model changes. Even if everything else remains the same, plugging in a different LLM into an existing system can produce regression faults, i.e., such a chance may break a functionality that was working well with the previous model. We note that the concept of such regression faults are generally lacking in current testing frameworks.

\subsection{Summary of Empirical Findings}

Across the three investigations, a clear pattern emerges. Current LLM testing tools (Sections \ref{sec:mapping} and \ref{sec:prompt}) have extensive support for atomic evaluations\slash oracles but provide little systematic support for aggregated oracles, reproducibility, or variability-aware assessment. Both the manual and LLM-assisted analyses reveal similar structural gaps: tools rarely distinguish between goals and properties, offer limited mechanisms for measuring input coverage, and focus on reproducibility rather than true stochastic variability. The LLM-based evaluation further confirmed that while LLM models themselves can efficiently surface relevant evidence when evaluating new testing tools, human interpretation remains essential for judging the depth and validity of tool capabilities.

The DroidAgent sensitivity study (Section \ref{sec:sensitivity}) illustrates these limitations in practice, showing how configuration changes and nondeterminism directly affect observed behavior in LLM-based systems.

Together, these insights establish a foundation for the discussion that follows which explores how the field---and the taxonomy itself---must likely evolve to address these challenges and to guide the development of more reliable, scalable, and adaptive LLM testing methodologies toward 2030.

\section{Discussion: Towards 2030}
\label{sec:2030}

The taxonomy proposed in this paper offers a conceptual foundation for reasoning about how to test systems built around or including large language models, while also revealing areas where current practice lags behind. Our analyses of existing tools and frameworks expose a mismatch between what is measured and what actually varies in such systems. \textbf{Most tools treat each test execution as an isolated event}, relying on atomic oracles that presume determinism. Yet as our sensitivity study and tool mapping show, variability---across models, configurations, and even repeated runs of the same prompt---is an intrinsic property of these systems. Treating this variability not as noise but as a design variable marks a critical shift: \textbf{correctness must be viewed as probabilistic rather than absolute}. This reframing moves testing from validating single outputs to characterizing behavioral distributions across repeated or structurally diverse executions.

Extending the taxonomy into a stepwise test design methodology could make this conceptual shift more practical. Such a method would guide practitioners from defining test goals to selecting suitable oracles and aggregation strategies. Aggregated oracles bridge deterministic testing and stochastic behavior by replacing binary pass\slash fail judgments with confidence estimates and distributions of outcomes. Traditional testing already employs implicit aggregation---regression suites track pass\slash fail rates over time, CI\slash CD pipelines monitor flakiness, and developers interpret trends across multiple runs. Our contribution is not to introduce aggregation as a new idea, but to argue that it should become a first-class concern in LLM testing frameworks. Whereas traditional tools aggregate results retrospectively and informally, LLM testing requires deliberate aggregation strategies that acknowledge model non-determinism rather than treating it as a defect. Future work could refine this distinction between \emph{temporal aggregation}, combining results across repeated runs, and \emph{structural aggregation}, combining results across models, prompts, or input variants. Existing research on probabilistic software properties can be revisited~\cite{grunske2008specification}, yet open challenges remain regarding optimal sample size, cost and accuracy trade-offs, and how to convey probabilistic findings clearly to human stakeholders.

The empirical findings further underscore the need for improved tool support. While frameworks such as DeepEval, Promptfoo, and Giskard begin to incorporate aspects of variability-aware testing, their metrics and datasets still reflect traditional, deterministic paradigms. A key opportunity lies in \textbf{developing evaluation ecosystems where humans and LLMs jointly serve as oracles}. In these hybrid settings, human judgment provides grounding, while LLM-based evaluators can contribute scale and adaptivity. Understanding how to calibrate, govern, and ensure transparency in such human-in-the-loop testing will be essential.
An important open question is how to design user interfaces and interaction patterns that enable domain experts and other stakeholders beyond traditional software testers to effectively participate in testing, as they possess the contextual knowledge necessary to validate LLM behavior in specialized domains. This shift may introduce new challenges in maintaining consistent evaluation standards and in training non-technical stakeholders to interpret probabilistic test outcomes effectively. Questions of bias, fairness, and accountability arise when automated judges themselves may vary in output and vary differently for different sets of inputs, reinforcing the need for auditability and explainability of probabilistic verdicts.

SUT variability, stemming from differences in model versions, configurations, and interaction contexts, adds another dimension of instability that current methods seldom capture. The DroidAgent empirical investigation illustrates how minor configuration changes can dramatically alter observed behavior, emphasizing the need for regression testing that spans both versions and parameterizations. Future work should explicitly model such drift to determine when and why system behavior degrades. Statistical methods such as early stopping~\cite{wald1948optimum} and testing techniques like property-based testing~\cite{claessen2011quickcheck} may help balance cost and confidence. Developing adaptive selection strategies that automatically adjust oracle types, sample sizes, and evaluation criteria based on observed variability would further enhance test automation, ensuring that test suites evolve alongside the systems they evaluate while minimizing testing costs.

Testing is increasingly entangled with development in LLM-based systems, representing a fundamental shift from traditional software engineering workflows. Where testing once served mainly as a post-development validation step, it can now play an important role throughout the entire lifecycle of LLM systems, from initial prompt engineering and model selection to iterative refinement and deployment. This integration shares similarities to test-driven development but operates under different constraints: tests must adapt as \textbf{models evolve, prompts are refined based on test outcomes, and the boundary between writing code and configuring behavior is more blurred}. Continuous testing paradigms, where evaluation is embedded into every stage of system evolution, may offer valuable insights for managing this entanglement. Understanding how to operationalize testing not merely as a gatekeeper but as a co-design mechanism will be critical for mature LLM engineering practices.

Although this work introduces a conceptual taxonomy rather than a full testing framework, it implicitly proposes a concrete research agenda. Key next steps include: formalizing the notion of \emph{probabilistic correctness} in the context of LLM testing, defining standard metrics for aggregated verdicts, and building toolkits that support hybrid human/LLM oracle workflows. Broader validation will also be necessary, through replication across domains, longitudinal tracking of variability over time, and empirical benchmarks that measure aggregated oracle performance under real-world conditions.

Beyond technical methodology, this shift toward probabilistic correctness has societal implications. As AI-based systems take on roles with higher stakes, accountability and fairness in their evaluation become more central concerns. Variability-aware testing frameworks could enhance transparency and trust, but they also demand clarity about who is responsible when outcomes are probabilistic rather than deterministic. Making these uncertainties explicit can be an important step toward more honest and explainable software systems where AI models are key components.

We argue that \textbf{progress toward mature LLM testing will depend on close collaboration between academia, industry, and policymakers}. Open-source ecosystems already provide a foundation for standardization, but maintaining their relevance requires shared benchmarks, consistent terminology, and transparent reporting of stochastic outcomes—efforts that demand coordinated engagement across all three sectors. Researchers can distill empirical insights into generalizable principles, while practitioners validate them in operational environments, and policymakers can help establish governance frameworks that ensure accountability and ethical deployment. Even if these more traditional roles seem to increasingly blur as AI systems become more deeply embedded in society, we argue they will all continue to be relevant.

By recognising ambiguity and variability as intrinsic properties rather than flaws, the community can move toward a more realistic and principled understanding of correctness in LLM-based systems. The proposed taxonomy can offer a shared vocabulary and some direction for this transition, one that connects traditional software testing with the emerging realities of probabilistic, adaptive, and human-interactive intelligent systems.

\bibliographystyle{ACM-Reference-Format}
\bibliography{lib}

\end{document}